\documentstyle[aps,preprint,eqsecnum]{revtex}
\tightenlines
\begin{document}
\draft
\title{Decay of Z-string due to the fermion emission.}
\author{A.A. Kozhevnikov}
\address{Laboratory of Theoretical Physics,\\
S.L. Sobolev Institute for Mathematics,\\
630090, Novosibirsk-90, Russian Federation
\footnote{Electronic address: kozhev@math.nsc.ru}
}
\date{\today}
\maketitle
\begin{abstract}

The question of classical topological stability of the gauge vortex defects
is reanalyzed upon taking into account the quantum perturbative effects.
The purely Abelian Higgs string remains stable
while the coupling of the effectively
Abelian Z-string (with the fixed zero upper component of the scalar doublet)
to the fermions of the minimal standard model is shown to result in its
decay, via the fermionic pair emission. The decay rate is scaled as $10^{-1}$
of the decay rate of $Z^0$ boson. The influence of the surrounding
charge-asymmetric fermionic matter is considered, demonstrating the dependence
of the decay rate on the contour shape and its suppression at sufficiently
large charge asymmetry.
\end{abstract}

\narrowtext

\section{Introduction}
\label{sec1}

The interest to the electroweak (EW) defects  in the form of Z-strings
\cite{ews} that could be produced at the phase transition in early
universe is attributed, in particular, to their possible role in the
processes of baryogenesis at the EW scale \cite{rubakov96}. First, they are
treated as  possible carriers of nonzero baryon number \cite{vachas94,barrio},
so that the change of this number
$\Delta B$ is related to the change of the Chern-Simons (helicity) number
$\Delta n_{\rm CS}$ of the Z-string configuration via the integrated anomaly
equation for the baryon number current \cite{thooft76}:
\begin{equation}
\Delta B=\Delta n_{\rm CS}=N_f\frac{\bar g^2}{(4\pi)^2}
\cos(2\theta_W)\Delta
\int d^3x{\bf Z}\cdot (\mbox{\boldmath$\nabla$}\times{\bf Z}),
\label{eq1}                                              
\end{equation}
provided the background electromagnetic field is absent. Here $g$ and
$g^\prime$ are, respectively, the SU(2) and U(1) coupling constants, $\bar g=
\sqrt{g^2+g^{\prime 2}}$, $\theta_W$ is the Weinberg angle, and $N_f$ is the
number of the fermionic families. The last integral in Eq.~(\ref{eq1}) is
recognized to be the helicity number $h_{\rm Z}$ of the EW Z-string
configuration. Second, EW strings were suggested to be the source
\cite{branden94} of nonequilibrium \cite{sakharov} required  in any model
of baryogenesis, to produce the baryon density via the evolution of the EW
string network.

Although the simplest variant of Z-string solution is known to be unstable
\cite{ews} towards to the development of either the upper component of the
Higgs field doublet \cite{james92,nacul}, or to $W$ condensation
\cite{perkins}, these instabilities can be possibly neutralized by particles
bound to Z-string \cite{nacul,watk} or by imposing external magnetic field
\cite{garriga}, respectively. The two Higgs doublet generalization of the
Minimal Standard Model (MSM) admits classically stable Z-string \cite{dvali}.

In the papers cited above the possible quantum properties of Z-strings
were neglected despite the fact that these defects possess the
microscopic transverse
size of the order of $\sim m^{-1}_Z$ and could behave similar to  $Z^0$ bosons
and, in particular, become unstable via the emission of fermions. The latter
point should be clarified from the very start. Indeed, the
Z-string with the fixed zero upper component of the Higgs field doublet can be
treated as an effectively Abelian Abrikosov-Nielsen-Olesen (ANO) string
\cite{ano}  in the Higgs model with the action
\begin{equation}
S=\int d^4x\left[-{1\over4}F^2_{\mu\nu}+|(\partial_\mu+igA_\mu)\phi|^2
\right.
-\left.{\lambda^2\over2}\left(|\phi|^2-{\eta^2\over2}\right)^2\right].
\label{action}
\end{equation}
This string is known to be topologically stable. The formal proof is reduced
to the statement that the homotopy
$f(\rho)[(1-u)\exp(i\theta)+u]$ interpolating between the two {\it static}
configurations of the scalar field $\phi$ with different winding numbers
$n=1$ at $u=0$ and $n=0$ at $u=1$, is not admissible since the energy barrier
$\Delta E\propto u^2(1-u)^2\int_0^\infty d\rho\rho f^4(\rho)$,
separating these two configurations with the finite energy per unit length
is infinite. Here $g$ is the U(1) coupling constant, the profile of the scalar
field $f(\rho)$ has the asymptotic  $f\to\eta/\sqrt{2}$ at $\rho\to\infty$,
$\theta$ is the polar angle in the  $xy$ plane, $\rho=\sqrt{x^2+y^2}$.
In quantum case one  examines the path integral
\begin{equation}
\langle\Psi_f|\Psi_i\rangle=\int d[\phi_f]\Psi_f^\ast[\phi_f]
\int_{\phi(t_i)=\phi_i}^{\phi(t_f)=\phi_f}d[\phi]
\exp(iS[\phi])\Psi_i[\phi_i]d[\phi_i],
\label{path}
\end{equation}
($\phi$ is the shorthand notation for all relevant fields)
which determines the overlap of two wave functionals. The latter is claimed
to be vanishing, if the fields $\phi_i$ and $\phi_f$ belong to distinct
homotopical classes \cite{finkel} characterized in the present case by
different winding numbers of the scalar field, since the infinite energy
barrier makes the exponent to be rapidly oscillating. How is evaded this
topological veto in the course of particle emission?

To this end one should emphasize the following. First, the above homotopy
determines the field configuration that does not obey the equations of motion,
except for the border values $u=0$ and $u=1$,
provided $u$ is taken as dynamical
variable. Admitting it as the quantum trajectory which does not oblige to
obey the classical equations of motion, one encounters the infinite energy
barrier making the amplitude vanishing. So, strictly speaking, only this
specific type of the field deformation is not allowed quantum mechanically.
On the other hand, the time dependent winding number $n(t)$
could interpolate between different topologies without going beyond the
class of fields with finite energy per unit length,
$\phi=\eta\exp[in(t)\theta]/\sqrt{2}$. This is still the solution of the
equations of motion, provided the time component of the vector potential
is induced via the condition of the vanishing covariant time derivative
of the scalar field. (See Sec. \ref{sec2} below).
Classically, however, the winding number
being the degree of mapping should be an integer which would seem to leave
only the known possibility of the conserved $n$.
One of the purposes  of the present
paper is to point out that this conclusion can be evaded
by treating this number as quantum dynamical variable $n\equiv n(t)$.
To this end we go beyond the approximation of the classical background
field and quantize the vortex field configuration. The latter,
in the approximation of the Higgs boson mass being much greater than the gauge
boson mass called sometimes as the London limit,
is known to be characterized solely by the spacetime dependent
phase of the scalar field.
The demand of the scalar field to be  single valued
should be imposed then only on the classical field configuration. Precisely,
the expectation value of $n$ in the quantum state $\Psi$ should be an
integer. The wave function of this state $\Psi$ will be found explicitly.
The next step is to study the evolution of this wave function under the
influence of the interaction of the vortex background with the propagating
excitations. It will be shown that the wave function in its dependence
on time and the winding number acquires the components peaked at successively
diminishing winding numbers which is interpreted as violation of the
topological stability of the classical string
configuration. The lifetime of the string with the unit winding number turns
out to be proportional to  the lifetime of the gauge boson of the model, with
the factor depending on the string contour.
The purely Abelian string in the model (\ref{action}) survives this mechanism
due to kinematical reason, since the net mass of the final particles
entering the vertex of their interaction  with the string background
is greater than the energy splitting of the quantum levels
between which the transition occurs followed by the particle emission.
The emergence of the levels results from treating the winding number as
quantum variable \cite{kozhev95a}. It should be emphasized that  this
type of instability is by no means include the transition via the infinite
energy barrier mentioned above.

	In order to fix the notations,
let us consider the neutral current piece of the lagrangian density
of MSM assuming for a while a single fermionic family:
$${\cal L}_{\rm NC}={\cal L}_{\rm boson}+{\cal L}_{\rm fermion}+
{\cal L}_{\rm Yukawa},$$
where the bosonic part is
\begin{equation}
{\cal L}_{\rm boson}=-\frac{1}{4}Z^2_{\mu\nu}+|(\partial_\mu+
\frac{i}{2}\bar g
Z_\mu)\Phi|^2-\frac{1}{2}\lambda^2(|\Phi|^2-\frac{1}{2}\eta^2)^2,
\label{eq2}                                                   
\end{equation}
and the upper component of the Higgs doublet is taken to be zero; the
fermionic part is
\begin{eqnarray}
{\cal L}_{\rm fermion}&=&i(\bar\psi_{+L},\bar\psi_{-L})
\left[\widehat{\partial}-i\bar g(T_3-Q_L\xi)\widehat{Z}\right]
\left(\psi_{+L}\atop\psi_{-L}\right)+             \nonumber\\
& &i\bar\psi_{+R}\left[\widehat{\partial}-
i\bar g(-Q_{+R}\xi)\widehat{Z}\right]\psi_{+R}+
i\bar\psi_{-R}\left[\widehat{\partial}-
i\bar g(-Q_{-R}\xi)\widehat{Z}\right]\psi_{-R};
\label{eq3}                                                     
\end{eqnarray}
while
\begin{equation}
{\cal L}_{\rm Yukawa}=-h_-(\bar\psi_{-L}\psi_{-R}\Phi+\mbox{c.c.})-
h_+(\bar\psi_{+L}\psi_{+R}\Phi^{\ast}+\mbox{c.c.}).
\label{eq4}                                                      
\end{equation}
gives the mass to the fermions and describes their interaction with the Higgs
boson. Note that $h_+=0$ in the case of leptons.
Hereafter $\widehat{\partial}\equiv\gamma^\mu\partial_\mu$ etc.,
$Z_{\mu\nu}=\partial_\mu Z_\nu-\partial_\nu Z_\mu$, $\xi=\sin^2\theta_W$,
$T_3$ is the third component of the weak SU(2) isospin,
and $Q_L=\mbox{diag}(Q_{+L},Q_{-L})$ is the charge matrix of the left
fermions, $Q_{\pm L(R)}$ being corresponding electric charge.

The subsequent material is organized as follows. Section \ref{sec2} contains
the derivation of the effective reduced action for the gauge vortex state in
terms of the winding number, including the discussion of the contour motion
omitted in Ref. \cite{kozhev95a}. It is  argued there why the dynamics of the
radial part  (modulus) of the scalar field is inessential for evaluation of
the lifetime of the string in the London limit.
The quantum state of the string configuration and the
criterion of its stability against the
particle emission is discussed in Sec. \ref{sec3} . The purely Abelian
Higgs model admits the string stable against this mechanism. This is not the
case for effectively Abelian Z-string which possesses the coupling with
almost massless fermions. The calculation of the corresponding
decay rate of the state of Z-string having arbitrary geometric shape, including
the parity-odd contours, is presented in Sec. \ref{sec4}, with the taking
into account of the effects of the nonzero fermionic density. The damping of
the energy, Z-flux and helicity (Chern-Simons) number of Z-string is discussed
in Sec. \ref{sec5}.
Sec. \ref{sec6} is devoted to the discussion of the validity of
assumptions adopted in the paper, and to the conclusion drawn from the
present study.

\section{The  action of Z-string in terms of the winding number}
\label{sec2}

The fact that the Z-string solution \cite{ews}
is an embedding into SU(2)$\times$U(1) gauge
group of the known U(1) ANO vortex solution \cite{ano} [compare Eqs.
(\ref{action}) and (\ref{eq2})] permits one
to refer to our earlier derivation \cite{kozhev95a} of the reduced effective
action. When so doing, it is very comfortable to use the space Fourier
transforms of the fields, which are elementary functions rather than the
special ones used in writing down the original solution \cite{ews}.

The nonstationary field configuration of Z-string is expressed through  the
spacetime dependent phase $\chi\equiv\chi({\bf x},t)$ of the
Higgs field $\Phi({\bf x},t)=\eta\exp(i\chi)/\sqrt{2}$
in the London limit.
This is the limit of $m_H\gg m_Z$, $\ln m_H/m_Z$ is also large, where
$m_Z=\bar g\eta/2$ and $m_H=\lambda\eta$ are the
masses of the $Z^0$ and Higgs bosons;  $\lambda$ and
$\eta/\sqrt{2}$
are the Higgs field self-coupling and magnitude.
Let us remind briefly why the dynamics of the phase $\chi$ of the scalar
field, not the dynamics of its radial part (modulus)
$f$, is the only one essential in this limit.
To this end one should rewrite the action Eq. (\ref{action}) in terms of these
variables,
\begin{eqnarray}
S&=&\int d^4x\left\{{1\over2}(-\partial_t{\bf A}-
\mbox{\boldmath$\nabla$}A_t)^2-{1\over2}(\mbox{\boldmath$\nabla$}
\times{\bf A})^2+f^2[(\dot\chi+gA_t)^2-(\mbox{\boldmath$\nabla$}\chi-
g{\bf A})^2]\right.     \nonumber\\
&&\left.+\dot f^2-(\mbox{\boldmath$\nabla$}f)^2
-{\lambda^2\over2}\left(|\phi|^2-{\eta^2\over2}\right)^2
\right\},
\label{act1}
\end{eqnarray}
and to convince  that the contribution of the radial part to the
path integral Eq. (\ref{path}) is factored out in the London limit.
Indeed, an exact factorization could be broken, in principle, in the
following two cases. First, there is the mixed third term in the first line of
Eq. (\ref{act1}). But for large distances  $\rho\agt m^{-1}_H$
one may set $f\simeq\eta/\sqrt{2}$, which leaves only the phase, while at the
short distances  $\rho< m^{-1}_H$ the amplitude of the Higgs condensate
behaves as  $f\sim\rho^n$ \cite{ano}, so this term gives negligible
contribution. The conclusion remain valid in the case of oscillating $n$
(see below), since one should average the contribution over the period of
oscillations after which it is not the winding number itself but its amplitude
that enters into expression leaving the suppression at $\rho\to0$ intact.
Second, the term  $\dot f^2$ is proportional to
$\dot n^2$ at the short distances, however, its contribution to the action
is suppressed as $({\eta\over m_H})^2$ as compared to the logarithmically
enhanced ($\ln{m_H\over m_Z}\gg1$) contribution of large distances
$\rho\agt m^{-1}_H$ coming from the terms containing the phase.
So, one can ignore the details of the Higgs field profile $f$, taking it to be
uniform $\eta/\sqrt{2}$ in all coordinate space except the vortex line where
it approaches zero at characteristic distances $\sim m_H^{-1}$. The
contribution of the radial part is then factored out as redefinition of the
measure in the path integration over $n(t)$.

The equation for the Z-magnetic field,
\begin{equation}
\mbox{\boldmath$\nabla$}\times{\bf H}_Z=\frac{2m^2_Z}{\bar g}
\mbox{\boldmath$\nabla$}\chi-m^2_Z{\bf Z},
\label{eq22}
\end{equation}
is solved to give the Z-magnetic field strength,
\begin{equation}
{\bf H}_Z({\bf k},t)=\frac{4\pi n_a}{\bar g}\cdot
\frac{m^2_Z}{{\bf k}^2+m^2_Z}\oint d\sigma{\bf X}^\prime_a
\exp(-i{\bf k\cdot X}_a),
\label{eq23}
\end{equation}
and the vector potential:
\begin{equation}
{\bf Z}({\bf k},t)=\frac{4\pi n_a}{\bar g}\cdot
\Biggl(\frac{1}{{\bf k}^2}-\frac{1}{{\bf k}^2+m^2_Z}\Biggr)
\oint d\sigma i[{\bf k}\times{\bf X}^\prime_a]
\exp(-i{\bf k\cdot X}_a).
\label{eq23a}
\end{equation}
Here the integral over $\sigma$ comes from the equation for the phase $\chi$
read off from Ref.~\cite{singph}, with the proper
continuation to the Minkowski spacetime:
\begin{equation}
\mbox{\boldmath$\nabla$}\times\mbox{\boldmath$\nabla$}
\chi({\bf x},t)=2\pi n_a\oint d\sigma
{\bf X}_a^{\prime}\delta^{(3)}
[{\bf x}-{\bf X}_a(\sigma,t)],
\label{eq8}                                                    
\end{equation}
where ${\bf X}_a\equiv{\bf X}_a(\sigma,t)$ is the evolving closed
string contour $a$ parametrized by the arclength $\sigma$.
 Hereafter the prime over ${\bf X}$ will
denote the derivative with respect to corresponding parameter along the
contour, while
the overdot will do the time derivative. The case of many contours is embraced
by taking the sum over individual  contributions in the right hand side of
Eq.~(\ref{eq8}). It is argued in Ref. \cite{earnsh94} that
possible fermionic zero modes \cite{jackros81}  do not perturb the string
profile.
Recall that the winding number  $n_a\equiv n_a(t)$ of the scalar field
is directly related with the  number of quanta of the magnetic-like flux,
in the present case the Z-flux, via the condition of the vanishing
covariant derivative of the Higgs field deep inside in the Higgs condensate,
\begin{equation}
\oint{\bf Z}\cdot d{\bf l}={2\over\bar g}\oint
\mbox{\boldmath$\nabla$}\chi\cdot d{\bf l}=\phi_0n_a,
\label{match}
\end{equation}
where $\phi_0={4\pi\over\bar g}$ is the quantum of Z-flux.
It is essential that the first term in the parentheses of Eq. (\ref{eq23a})
is in fact a pure gauge one. Indeed, the Fourier component of
$${\bf v}({\bf x},t)\equiv\frac{2}{\bar g}\mbox{\boldmath$\nabla$}\chi$$
found from Eq. (\ref{eq8}), is
\begin{equation}
{\bf v}({\bf k},t)=\frac{4\pi n_a}{\bar g{\bf k}^2}
\oint d\sigma i[{\bf k}\times{\bf X}^\prime_a]
\exp(-i{\bf k\cdot X}_a).
\label{eq25}
\end{equation}
Locally, in a plane transverse to the tangent vector ${\bf X}^\prime_a$,
Eq. (\ref{eq25}), after going back to the coordinate space, leads to the
transverse components of ${\bf v}$ to be
$$v_i({\bf x},t)\propto\varepsilon_{ij3}\partial_j\ln
|\mbox{\boldmath$\rho$}|=\partial_i
\mbox{atan}\rho_2/\rho_1\mbox{; }{\bf X}^\prime_a\cdot\mbox{\boldmath$\rho$}
=0.$$

To specify the dynamical part of the problem, one should write down the
Z-electric field ${\bf E}_Z=-\mbox{\boldmath$\nabla$}Z_t-\partial_t{\bf Z}$,
where
\begin{equation}
Z_t=-\frac{2}{\bar g}\partial_t\chi
\label{eq26}
\end{equation}
replaces the  condition $Z_t=0$ appropriate in the static case.
One has
\begin{equation}
{\bf E}_Z=\frac{2}{\bar g}\mbox{\boldmath$\nabla$}\partial_t\chi-
\partial_t{\bf Z}=\frac{2}{\bar g}(\mbox{\boldmath$\nabla$}\partial_t-
\partial_t\mbox{\boldmath$\nabla$})\chi+\partial_t\Biggl(
\frac{2}{\bar g}\mbox{\boldmath$\nabla$}\chi-{\bf Z}\Biggr).
\label{eq27}
\end{equation}
The commutator of the derivatives  is nonzero in view of the
singular character of the phase $\chi$ \cite{singph},
so the Fourier component of ${\bf E}_Z$ becomes
\begin{eqnarray}
{\bf E}_Z({\bf k},t)&=&-{4\pi\over\bar g}n_a\oint d\sigma
(\dot{\bf X}_a\times{\bf X}_a^{\prime})\exp
[-i{\bf k}\cdot{\bf X}_a(\sigma,t)]    \nonumber\\
& &+\frac{{\bf k}^2}{{\bf k}^2+m^2_Z}
\partial_t{\bf v}({\bf k},t).
\label{dop1}
\end{eqnarray}

Note that the vector potential ${\bf Z}$ and the magnetic ${\bf H}_Z$ type
field strength  can also be expressed through the gradient of the
singular phase $\chi$  as
\begin{eqnarray}
{\bf Z}({\bf k},t)&=&
\left(1-\frac{{\bf k}^2}{{\bf k}^2+m^2_Z}\right){\bf v}({\bf k},t),
      \nonumber\\
{\bf H}_Z({\bf k},t)&=&{m_Z^2\over{\bf k}^2+m^2_Z}
i[{\bf k}\times{\bf v}({\bf k},t)].
\label{fields}
\end{eqnarray}
In what follows we will omit the encounters of the nearby string segments.
Important are in the processes of the string rearrangements, they cannot
be described in the framework of the London limit and demands the numerical
integrations of the full set of the equations of motion.
Substituting Eqs. (\ref{eq23}), (\ref{eq23a}) and (\ref{dop1}) into the
lagrangian $L_{\rm boson}=\int d^3x{\cal L}_{\rm boson}$ one obtains,
with the help of the relation
$$\int d^3x{\bf H}_Z^2({\bf x})=\int d^3k|{\bf H}_Z({\bf k})|^2
/(2\pi)^3$$ etc,
the expression for the action of the single gauge vortex:
\begin{eqnarray}
S_{\rm vortex}&=&
{\phi_0^2\over2(2\pi)^3}\int{d^3k\over({\bf k}^2+m^2_Z)^2}
\int dt\oint d\sigma_1d\sigma_2\exp\{i{\bf k}\cdot[{\bf X}(\sigma_1)
-{\bf X}(\sigma_2)]\}           \nonumber\\
&& \times\{[\dot n_a^2 {\bf k}^2-m^2_Z({\bf k}^2+m^2_Z)n_a^2]
[{\bf X}^\prime(\sigma_1)\cdot{\bf X}^\prime(\sigma_2)]   \nonumber\\
& & +n_a^2({\bf k}^2+2m^2_Z)
({\bf k}\cdot[\dot{\bf X}(\sigma_1)\times{\bf X}^\prime(\sigma_1)])
\cdot({\bf k}\cdot[\dot{\bf X}(\sigma_2)\times{\bf X}^\prime(\sigma_2)])
                \nonumber\\
& &+n_a^2m^4_Z[\dot{\bf X}(\sigma_1)\times{\bf X}^\prime(\sigma_1)]
[\dot{\bf X}(\sigma_2)\times{\bf X}^\prime(\sigma_2)]\}.
\label{lagr}
\end{eqnarray}
Let us make a step apart and show with the method similar to those of
P.~Orland, Ref. \cite{singph} and \cite{sato95} how the known Nambu-Goto (NG)
action results from Eq. (\ref{lagr}). To this end one should take the limit
of the fixed winding number $n_a$ and set the Z-boson  mass $m_Z\to\infty$
before the momentum integration. Then the term $\propto\dot n_a$ drops out,
and the action becomes, in the gauge $X^0\equiv t=\tau$,
\begin{eqnarray}
S_{\rm NG}&=&{\phi^2_0\over2}\int d^2s_1d^2s_2\delta^{(4)}[X(s_1)-
X(s_2)]\{-{\bf X^\prime}(s_1)\cdot{\bf X^\prime}(s_2)
\nonumber\\
& &+ [{\bf\dot X}(s_1)\times{\bf X^\prime}(s_1)]\cdot
[{\bf\dot X}(s_2)\times{\bf X^\prime}(s_2)]\},
\label{ng1}
\end{eqnarray}
where  $s_{1,2}\equiv s^A_{1,2}=(\tau_{1,2},\sigma_{1,2})$ is the two-
dimensional vector. Using the Gaussian regularization
of the $\delta$ function and the expansion
\begin{equation}
{\bf X}(s_2)\simeq{\bf X}(s_1)+(s_2-s_1)^A\partial_A
{\bf X}
\label{eqcl}
\end{equation}
\cite {sato95} valid under the condition
$|{\bf X^{\prime\prime}}(\sigma)|\ll m_Z$, one obtains
\begin{eqnarray}
S_{\rm NG}&=&{1\over2}\Biggl({\phi_0\over2\pi\Lambda^2}\Biggr)^2
\int d^2s_1d^2z\exp(-{1\over2\Lambda^2}z^Az^B\partial_AX^\mu
\partial_BX_\mu)(-{\bf X^\prime}^2+[{\bf\dot X}\times{\bf X^\prime}]^2)
             \nonumber\\
& &={\phi^2_0\over4\pi\Lambda^2}\int d^2s\sqrt{\mbox{det}
\partial_AX^\mu\partial_BX_\mu},
\label{ng}
\end{eqnarray}
where $\Lambda^{-1}\to\infty$ is an ultraviolet cutoff, $\partial_A=
\partial/\partial z^A$ and $\mbox{det}\partial_AX^\mu\partial_BX_\mu=
-{\bf X^\prime}^2+[{\bf\dot X}\times{\bf X^\prime}]^2$ in the chosen gauge.
Up to an overall factor, the last equality in Eq. (\ref{ng}) is recognized
to be the NG-action.

Coming back to the case of large but finite $m_Z$, one obtains
the action as the sum
$\sum_a S^{(a)}_{\rm vortex}$, where
\begin{equation}
S^{(a)}_{\rm vortex}=
{4\pi\over\bar g^2}\ln{m_H\over m_Z}\int dt\oint d\sigma
\left\{(\dot n_a^2-m^2_Zn_a^2){\bf X}^{\prime2}_a
+m^2_Zn^2_a
[{\bf\dot X}_a\times{\bf X}^\prime_a]^2/c^2_0\right\}.
\label{acti}
\end{equation}
Here the first and the third terms come from the electric-type field, while
the second term comes from the kinetic energy of the scalar field, thus
demonstrating an intimate interplay of the vector and scalar fields in the
string solution. The energy of Z-magnetic field is not enhanced
logarithmically in the London limit and by this reason is dropped. Further,
$$c^2_0={4m^2_Z\over O(1)m^2_H}\ln{m_H\over m_Z}\ll1$$
is the velocity squared which characterizes the classical string motion in
the case of finite masses,
the factor $O(1)$ reflects the ignorance of the true Higgs field profile,
and $m_H$ appears as the natural upper limit of the integration over momentum.
We retain only the terms in the action
that refer to the string background. Omitted
are the terms which correspond to the propagating  Z and Higgs bosons. They
result in the renormalization of the parameters of the action Eq. (\ref{acti})
and do not contribute, due to the energy conservation,
to its imaginary part (see below), which will be further of our main concern.
Eq. (\ref{acti}) is valid in the case of large but finite  $m_Z$,
so the terms corresponding to the interaction are exponentially small,
and only nearby segments of the string contour give an appreciable
contribution to the integral over arclength.

\section{Quantum state of the string configuration and the criterion of its
stability against the particle emission}
\label{sec3}

It is seen that under the condition $|{\bf\dot X}_a|\ll c_0$ assumed
hereafter, the dynamics of the winding number becomes
decoupled from the contour dynamics and is governed by the oscillator-like
action,
\begin{equation}
S^{(a)}_{\rm vortex}={1\over2}\int dt M_{\rm v}(\dot n^2_a-\omega^2n^2_a),
\label{dop12}
\end{equation}
where the  frequency and effective mass, in the gauge
${\bf X}^{\prime2}=1$, are, respectively,
$\omega=m_Z$ and
\begin{equation}
M_{\rm v}={8\pi L\over\bar g^2}\ln{m_H\over m_Z}.
\label{mass}
\end{equation}
Hereafter $L=\oint d\sigma$ is the length of Z-string in the chosen gauge.
The quantization subjected to the constraint
$\langle\psi|n|\psi\rangle=n_0$, where $n_0$ is an integer, is performed
with the help of the indefinite Lagrange multipliers and
gives the energy levels of the system
\begin{equation}
E_N=\varepsilon_{\rm v}Ln^2_0+m_Z(N+{1\over2}),
\label{levels}
\end{equation}
where $\varepsilon_{\rm v}=4\pi m^2_Z\bar g^{-2}\ln m_H/m_Z$
is the energy  per unit length of the vortex with  the unit winding number,
$N=0,1\cdots$ (do not confuse with the winding number  $n$)
\cite{fn1}.
The wave functions are the oscillatory ones,
$\psi_N(n)=\psi^{({\rm osc})}_N(n-n_0)$,
\begin{equation}
\psi^{({\rm osc})}_N(n-n_0)=\left({2\varepsilon_{\rm v}L\over\pi m_Z}
\right)^{1/4}{1\over\sqrt{2^NN!}}\exp\left[-{\varepsilon_{\rm v}L\over m_Z}
(n-n_0)^2\right]H_N\left[\sqrt{2{\varepsilon_{\rm v}L\over m_Z}}(n-n_0)
\right],
\label{wf}
\end{equation}
($H_N$ is the Hermit polynomial) displaced to $n_0$. They are sharply peaked
at the integer numbers.
Specifically, the wave function of the vacuum state without string
is $\psi^{(\rm osc)}_0(n)$, irrespective of the string contour
${\bf X}$ and the radial part of the scalar field.

Now one can evaluate the overlap integral (\ref{path}) for the string with
different winding numbers, $n_i$ at the moment $t=0$ and
$n_f$ at some later moment $t>0$. The string will be assumed to be in the
ground states, $N=0$, so that  its wave functions at those moments are
$\psi_i=\psi^{({\rm osc})}_0(n-n_i)$
and $\psi_f=\psi^{({\rm osc})}_0(n-n_f)$, respectively.
The result  depends crucially on the spectrum of the propagating
excitations, since the string interaction with the latter may result in an
imaginary correction to the energy levels Eq. (\ref{levels}).
As it will be shown below, such a correction emerges as the imaginary part
of the frequency of the $n$-oscillator,  $m_ZN\to (m_Z-i\Gamma/2)N$.
Using the known expression for the propagator of the damped oscillator, one
obtains the probability amplitude of the transition with the change of the
winding number:
\begin{equation}
\langle f|i\rangle=\exp\left\{-{\varepsilon_{\rm v}L\over2m_Z}
\left[n^2_i+n^2_f
-2n_in_f\exp(-\Gamma t/2-im_Zt)\right]\right\}.
\label{ampw}
\end{equation}
The expression for the probability averaged over the period
$T=2\pi m^{-1}_Z$ looks as
\begin{eqnarray}
\langle w_{fi}\rangle&=&{\cal N}
\exp\left[-{\varepsilon_{\rm v}L\over m_Z}(n^2_i+n^2_f)
\right] \mbox{I}_0\left[{2\varepsilon_{\rm v}L\over m_Z}n_in_f
\exp(-\Gamma t/2)\right]       \nonumber\\
&&\simeq{\cal N}\exp\left\{-{\varepsilon_{\rm v}L\over m_Z}\left[n_f-n_i
\exp(-\Gamma t/2)\right]^2\right\}
\label{prob1}
\end{eqnarray}
[I$_0(z)$ is the modified Bessel function of order zero],
where the contribution of the path integral over the radial part of the
scalar field, as is argued in Sec. \ref{sec2}, is taken into account in the
form of the constant normalization factor ${\cal N}$.

If the damping were absent, $\Gamma=0$, the expression for the relative
probability would have a sharp maximum at $n_i=n_f$,
thus reproducing at the quantum level the
topological conservation of the string winding number.
On the contrary, at $\Gamma\not=0$ the maximum is achieved
at the configurations with the decaying winding number. The character of the
decay is established by reconciling the integers $n_i$ and $n_f$ with the
continuous time dependence in Eq. (\ref{prob1}).
Indeed, somewhere within the time interval
$\Delta t$ from the start, such that $-2\Gamma^{-1}\ln(1-1/2n_i)<\Delta t
<-2\Gamma^{-1}\ln(1-3/2n_i)$, the nearest integer $n_f$ for which the
probability is nonzero, is $n_f=n_i-1$, and so forth down to $n_i=1$.
In particular,
the lifetime of the U(1) gauge string with $n=1$ is $2\Gamma^{-1}$.
The classical topological veto is evaded, because due to the damping the wave
functional of the initial string configuration acquires the
components of the vacuum state. Note that the topology change is
faster for initially multiple winding numbers, since the time of awaiting
of $n=n_i-1$ is
$\Delta t\sim1/\Gamma n_i\ll\Gamma^{-1}$ \cite{fn2},
so that the total decay time
is
$T\simeq2\Gamma^{-1}\sum_{k=1}^{n_i}k^{-1}\sim2\Gamma^{-1}
\ln n_i$. The last approximate expression has the same form as if the winding
number would change continuously in the course of the decay.

Let us confront the result of Eqs. (\ref{ampw}) and (\ref{prob1}) with
the usual approach to the string stability. What is implied in the present
approach is the preparation of the string in the quantum state in which
the expectation value of the winding number is kept fixed. It is
perfectly the case that takes place for the Abrikosov string in type II
superconductors immersed into external magnetic field, while no specific
mechanism has not yet been devised
for the Z-string. Now  remove the constrain sustaining the string.
Then the state with the fixed winding number becomes the superposition of
the stationary states of an unconstrained oscillator, so that the amplitude
of finding it in the state with different expectation value $n_f$
of the winding number at a later moment is given by Eq. (\ref{ampw}).
The feature  of the present approach is thus the  quantum mechanical
treatment of the background field configuration. It should be recalled in
this respect that the notion of the classical background field is itself
an approximation aimed to represent the condensate of indefinite number
of quanta of corresponding quantum field. One of the possible representatives
of such a condensate is the coherent state, in which the expectation value of
the field operator, in the present case the phase of the Higgs field, is
nonzero. The wave function $\psi^{({\rm osc})}_0(n-n_0)$ used in the above
calculation is just the wave function of the coherent state in which the
winding number of the scalar field is the integer $n_0$.
In this sense the usual topological stability of the classical configuration
would correspond to the permanently constrained string resulting in the
amplitude $\langle f|i\rangle\propto\delta_{n_in_f}$ irrespective of
the damping rate $\Gamma$. The r\"ole of the damping in this case is
reduced to relaxing the field configuration to that determined by external
conditions.

To establish if $\Gamma$ is zero or not and hence the condition of the string
stability, one should examine the spectrum of the propagating excitations and
their couplings with the string background. Expanding the action of purely
Abelian Higgs model Eq. (\ref{action}) into
the contributions of the string background and the excitations, which are
the massive neutral Higgs $\varphi_H$ and vector $A$  bosons, one can show
that the only relevant coupling is the  string$\to\varphi_H+A$ one. Since the
emission of the excitations occurs in the course of the transitions between
the levels spaced, with the evident replacement, by $\omega\simeq m_A$, its
contribution to $\Gamma$ is
forbidden by the energy conservation. Thus, the stability of the Abelian Higgs
string in the model Eq. (\ref{action}) with the neutral scalar field is merely
the kinematical consequence of rather limited spectrum of excitations 
\cite{instan}. It well may be not the case in other models.
In particular, the effectively Abelian Z-string with the fixed upper
component of the Higgs doublet illustrates this since possesses the
interaction with almost massless fermions.

\section{Evaluation of the decay width of Z-string.}
\label{sec4}

The fermionic loop correction to the action
Eq. (\ref{dop12}) can, in principle,
be evaluated by the integrating out the propagating fermions from the total
action of the standard model. However, we need only the imaginary part of the
resulting effective action. Since this
imaginary part arises due to real intermediate states on the mass shell,
one can use the unitarity relation for its evaluation
and, in turn,  Im$\omega$, allowing for the quantum transitions between the
energy levels of the oscillator Eq. (\ref{dop12}).

In order to obtain the  lagrangian of interaction of Z-string
with the physical fermions, one should rotate away the phase $\chi$ from
the mass term of the Lagrangian of the standard model Eq. (\ref{eq4}),
with zero upper component of the Higgs field. It can
be accomplished by the phase rotation of the chiral fermions,
\begin{eqnarray}
\psi_L&\to&\psi_L\exp[-2i\chi(T_3-Q_L\xi)],   \nonumber\\
\psi_{\pm R}&\to&\psi_{\pm R}\exp\left(2i\chi Q_{\pm R}\xi\right),
\label{eq14}                                                    
\end{eqnarray}
with the phases proportional to their respective Z charges,
$$Q_Z=T_3-Q\sin^2\theta_W;$$
$Q$ and $T_3$  standing, respectively, for the electric charge and the third
component of weak isospin of the chiral fermion.
The pure gauge term of the background field ${\bf Z}$ presented as the
first term in the parentheses of the expression for ${\bf Z}$
in Eq.~(\ref{fields}) absorbs the gradient of the rotation phase thus
leading to no shift of the lagrangian.
The term describing the interaction of the single chiral fermion, say
the left one, with the string becomes
$${\cal L}_{\rm int}=\bar g\bar\psi_L\widehat{Z}^{(1)}
Q_Z\psi_L,$$
where the Fourier transform of the space components of the short range
piece of the background Z-field ${\bf Z}^{(1)}$ is given by the second term
in parentheses of expression for ${\bf Z}$ in Eq. (\ref{fields}), while
the time component is zero in the London limit. Note that taking the limit
$m_Z\to\infty$ results in the decoupling of Z-flux from external fermions.

The matrix element of the emission of a pair of chiral
fermions by Z-string in the course of the quantum transition between
the levels $N$ and $N-1$ shifted by $m_Z$ is
\begin{equation}
{\cal M}_N=
-2\pi\bar gQ_Z\delta(m_Z-\varepsilon_1-\varepsilon_2)
\bar u_L(p_1)\mbox{\boldmath$\gamma$}\cdot{\bf V}
v_L(p_2)\langle N-1|n_a|N\rangle,
\label{eq16}                                                   
\end{equation}
where the above ${\bf Z}^{(1)}$ is factored out as $n_a{\bf V}({\bf p}_1+
{\bf p}_2)$, and the recoil is evidently negligible.
 Here $\mbox{\boldmath$\gamma$}$ is the spatial Dirac matrix,
$p_i=(\varepsilon_i,{\bf p}_i)$ is the
four-momentum of the final fermion $i=1,2$, and
\begin{equation}
\langle N-1|n_a|N\rangle=(N/2M_{\rm v}m_Z)^{1/2},
\label{jump}
\end{equation}
as is evident from the oscillator-like character of the
action Eq. (\ref{dop12}). The following calculation of the width
$\lambda_N$ of the Nth level includes the effects
of the particle-antiparticle asymmetry in the surrounding fermionic matter
by means of  the Fermi blocking
factors in the expression for the density of final states. Inserting the
identity $1=\int d^3k\delta({\bf k}-{\bf p}_1-{\bf p}_2)$, one obtains
\widetext
\begin{eqnarray}
\lambda_N&=&{N\bar g^2Q_Z^2\over8M_{\rm v}m_Z(2\pi)^5}
\int d^3k{d^3p_1d^3p_2
\over\varepsilon_1\varepsilon_2}\delta(m_Z-
\varepsilon_1-\varepsilon_2)\delta({\bf k-p_1-p_2})
      \nonumber\\
& &\times\sum_{\rm spins}
|\bar u_L(p_1)\mbox{\boldmath$\gamma$}\cdot{\bf V}
v_L(p_2)|^2[1-f(\varepsilon_1-\mu)][1-f(\varepsilon_2+\mu)]
     \nonumber\\
& &={N\bar g^2Q_Z^2\over16\pi M_{\rm v}m_Z}
\int{d^3k\over(3\pi)^3|{\bf k}|}\theta(m_Z-|{\bf k}|)
\int_{(m_Z-|{\bf k}|)/2}^{(m_Z+|{\bf k}|)/2}d\varepsilon\left\{
|{\bf V}|^2({\bf k}^2-m^2_Z+4m_Z\varepsilon)\right.   \nonumber\\
& &\left.+({\bf k}^2-m^2_Z)(2\varepsilon-m_Z)
i{\bf k}\cdot[{\bf V}\times{\bf V}^\ast]/{\bf k}^2
\right\}[1-f(\varepsilon-\mu)][1-f(m_Z-\varepsilon+\mu)].
\label{eq18}                                                  
\end{eqnarray}
\narrowtext
Here $T$, $\mu$ are, respectively, the temperature and chemical potential,
${\bf V}\equiv {\bf V}({\bf k})$,
$f(\omega)=(\exp{\omega\over T}+1)^{-1}$ is the Fermi distribution,
the fermions are taken massless, and the temperature dependence of
$m_Z$ is  assumed to be included.
The linear dependence of $\lambda_N=N\Gamma$
on $N$ means a nonzero imaginary part of the frequency
Im$\omega=-\Gamma/2$, hence the damping of the quantum
mechanical probability, with the rate $\Gamma$.

Internal integral over $\varepsilon$ in Eq. (\ref{eq18}) can be written as
\begin{eqnarray}
G(\mu,|{\bf k}|)&\equiv&
\int_{(m_Z-|{\bf k}|)/2}^{(m_Z+|{\bf k}|)/2}d\varepsilon\cdots=
|{\bf V}|^2({\bf k}^2+m^2_Z)I_1(|{\bf k}|)      \nonumber\\
& &+2\{2m_Z|{\bf V}|^2+({\bf k}^2-m^2_Z)
i{\bf k}\cdot[{\bf V}\times{\bf V}^\ast]/{\bf k}^2\}I_2(|{\bf k}|),
\label{integr}
\end{eqnarray}
where
\begin{eqnarray}
I_1(|{\bf k}|)&=&{T\over1-\exp(-m_Z/T)}\ln\frac{\mbox{ch}{\mu\over T}+
\mbox{ch}{m_Z+|{\bf k}|\over2T}}{\mbox{ch}{\mu\over T}+
\mbox{ch}{m_Z-|{\bf k}|\over2T}},      \nonumber\\
I_2(|{\bf k}|)&=&{1\over2}\mbox{sh}{\mu\over T}\exp{m_Z\over2T}
\int_{-|{\bf k}|)/2}^{(|{\bf k}|)/2}d\varepsilon
\frac{\varepsilon\mbox{sh}{\varepsilon\over T}}{\left(\mbox{ch}{m_Z\over2T}+
\mbox{ch}{\varepsilon-\mu\over T}\right)
\left(\mbox{ch}{m_Z\over2T}+
\mbox{ch}{\varepsilon+\mu\over T}\right)}.
\label{i12}
\end{eqnarray}
Because of ${\bf V}$ is rather complicated function of the momentum,
further evaluation of the decay width cannot be performed explicitly.
Instead, we will obtain useful approximate results valid in
the situations of the physical interest.

i) Empty space, $\mu=0$, $T=0$. The charge conjugation ($C$)-even,
parity ($P$)-odd, $CP$-odd structure
$i{\bf k}\cdot[{\bf V}\times{\bf V}^\ast]/{\bf k}^2$ drops after the
integration over final states as it should, since no $CP$ nonconserving
effects remain in the charge symmetric situation, in the absence of an
explicit $CP$ nonconservation in the lagrangian.
The necessary expression for $|{\bf V}({\bf k})|^2$,
$$|{\bf V}({\bf k})|^2=\left({4\pi\over\bar g}\right)^2
{{\bf k}^2\over({\bf k}^2+m^2_Z)^2}\oint d\sigma_1d\sigma_2
\left({\bf X^\prime_1}\cdot{\bf X^\prime_2}\right)
\exp[-i{\bf k}\cdot({\bf X_1}-{\bf X_2})],$$
where ${\bf X}_1\equiv{\bf X}(\sigma_1)$ etc, can be obtained for
sufficiently
smooth contours whose curvature satisfy the condition
$|{\bf X}^{\prime\prime}|/m_Z\ll1$. Then one can use the expansion
\begin{equation}
{\bf X}(\sigma_2)={\bf X}(\sigma_1)+{z\over1!}{\bf X}^\prime(\sigma_1)
+{z^2\over2!}{\bf X}^{\prime\prime}(\sigma_1)
+{z^3\over3!}{\bf X}^{\prime\prime\prime}(\sigma_1)\cdots,
\label{expan}
\end{equation}
where $z=\sigma_2-\sigma_1$,
to show that upon neglecting the terms with the second and higher
derivatives of the contour the following approximate expression holds:
\begin{equation}
|{\bf V}({\bf k})|^2\simeq2\pi\left({4\pi\over\bar g}\right)^2
{{\bf k}^2\over({\bf k}^2+m^2_Z)^2}
\oint d\sigma\delta({\bf k\cdot X^\prime}_a).
\label{eq18a}
\end{equation}
When obtaining Eq. (\ref{eq18a}), the integration over $z$ can be extended to
$\pm\infty$. Choosing the local (at given $\sigma$) coordinate system
${\bf k}=({\bf k}_\bot,{\bf k}\cdot{\bf X}^\prime)$, the ${\bf k}$
integration is easily performed.
The summation over all fermionic species, with  the expression
for the total width of $Z^0$-boson
$$\Gamma_Z=\bar g ^2m_Z\sum_{\rm fermions}Q_Z^2/24\pi,$$
gives the damping rate
\begin{equation}
\Gamma=\kappa(1-\ln2)\Gamma_Z.
\label{eq21}                                                   
\end{equation}
Hereafter the notation $$\kappa={3\over4\ln m_H/m_Z}$$ is used.
The length of the string $L$ drops from the final expression, since the
factor $L^{-1}$ coming from the probability of the quantum jump
$|\langle N-1|n_a|N\rangle|^2$ [see Eqs. (\ref{mass}) and (\ref{jump})],
is cancelled by another factor $L$
[see Eq. (\ref{eq18a})] arising
due to incoherent emission of a fermionic pair by arbitrary point along the
string.
Since $\ln{m_H/m_Z}$ is conjectured to be large, say 3, the damping
becomes about $10^{-1}$ of the decay rate of  $Z^0$ boson.
Note that the emission of the Higgs boson and of a pair of Z-bosons is
forbidden by the energy conservation, since in the lowest order adopted here
the energy released in the decay is $m_Z$. The photon emission occurs only
in higher orders, hence corresponding partial rate is suppressed.

ii) Cold, strongly charge-asymmetric fermionic matter,
$|\mu|/T\gg1$, $|\mu|>m_Z$. This is the case of the strong degeneracy of
the fermions, so one can write
\begin{equation}
[1-f(\varepsilon-\mu)][1-f(m_Z-\varepsilon+\mu)]\simeq\left\{
\begin{array}{c}
\exp(\varepsilon-\mu)/T\mbox{, } \mu>0\\
\exp(m_Z-\varepsilon-|\mu|)/T\mbox{, } \mu=-|\mu|<0.\\
\end{array}
\right.\;
\label{deger}
\end{equation}
Consider for the definiteness the case $\mu>0$. Then the expression for
$G(\mu,|{\bf k}|)$ reads
\begin{eqnarray}
G(\mu,|{\bf k}|)&\simeq&T\exp{m_Z/2-\mu\over T}\left\{|{\bf V}|^2\left[
\left(|{\bf k}|+m_Z\right)^2\exp{|{\bf k}|\over2T}-
\left(|{\bf k}|-m_Z\right)^2\exp{-|{\bf k}|\over2T}\right]\right. \nonumber\\
& &\left.+2|{\bf k}|\mbox{sh}{|{\bf k}|\over2T}\cdot({\bf k}^2-m_Z^2)
i{\bf k}\cdot[{\bf V}\times{\bf V}^\ast]/{\bf k}^2\right\}.
\label{gdeg}
\end{eqnarray}
Since the dominant contribution comes from $|{\bf k}|\sim m_Z$, and
$m_Z\gg T$, one may keep only the rising exponents in the calculation.
Furthermore, all the polynomial $|{\bf k}|$-dependent expressions in
the denominators of the ${\bf V}$-dependent expressions
can be approximated by their values at $|{\bf k}|=m_Z$. The
evaluation of the $CP$-odd contribution uses the expression
\begin{eqnarray*}
{i{\bf k}\cdot[{\bf V}\times{\bf V}^\ast]\over{\bf k}^2}
=\left({4\pi\over\bar g}\right)^2\oint d\sigma_1\oint d\sigma_2
{i{\bf k}\cdot[{\bf X}^\prime_1\times{\bf X}^\prime_2]
\over({\bf k}^2+m^2_Z)^2}\exp(-i{\bf k}\cdot{\bf X}_{12}),
\end{eqnarray*}
where ${\bf X}_{12}\equiv{\bf X}(\sigma_1)-{\bf X}(\sigma_2)$, together
with the dropping of the terms, which are suppressed by the powers of
$m_Z|{\bf X}_{12}|\gg1$. The integration over ${\bf k}$ and the inclusion
of the case $\mu<0$ results in the damping rate:
\begin{equation}
\Gamma\simeq\kappa\left({T\over m_Z}\right)^2
\langle\exp({m_Z-|\mu|\over T})\rangle\left(1+{TF_1\over m_Z}\right)
\Gamma_Z,
\label{colds}
\end{equation}
where
$$\langle(\cdots)\rangle=\sum_f(\cdots)Q^2_Z/\sum_fQ^2_Z,$$
is the average over Z-charges, and
\begin{equation}
F_1={8\over\pi L}\oint d\sigma_1d\sigma_2{{\bf X}_{12}
\cdot[{\bf X}_1^\prime\times{\bf X}_2^\prime]\over
{\bf X}_{12}^2}\sin m_Z|{\bf X}_{12}|
\label{f1}
\end{equation}
is the parity-odd factor which depends on the geometry of the string contour.

The damping is completely prohibited in this case. In fact, the energy of
emitted
fermion spreads from 0 to $m_Z$, so at $|\mu|>m_Z$ it necessarily hits the
filled states inaccessible by the Pauli exclusion principle. The
conditions of the complete suppression
are not likely to be achieved in the universe at large scale.
To be sure, the chemical potentials of {\it all} fermions should obey
the condition $|\mu|>m_Z$. But only for neutrinos there is a rather
high cosmological upper bound
$\left[\sum_\nu(\mu_\nu/T)^4\right]^{1/4}<45$ \cite{weinberg} for
the net contribution and considerably more
stringent bound from nucleosynthesis,
$|\mu_{\nu_e}|/T<0.2$, for the electronic neutrino \cite{kim}.
The baryonic and
leptonic asymmetries are small, $\sim10^{-8}-10^{-10}$. So the averaged
fermionic asymmetry of the universe is insufficient to prohibit the
decay strongly. Yet the smaller scale inhomogeneities in chemical potential
of all types of fermions at the EW epoch
could not be excluded by this argument.

iii) Cold, weakly charge-asymmetric fermionic matter, $|\mu|\ll T\ll m_Z$.
One has
\begin{eqnarray}
G(\mu,|{\bf k}|)&\simeq&|{\bf k}|\left\{|{\bf V}|^2
\left({\bf k}^2+m_Z^2+8\mu m_Z\exp{-m_Z\over2T}\right)\right.
 \nonumber\\
& &\left.+4\mu\exp{-m_Z\over2T}\cdot\mbox{ch}{|{\bf k}|\over2T}
\cdot({\bf k}^2-m_Z^2)
i{\bf k}\cdot[{\bf V}\times{\bf V}^\ast]/{\bf k}^2\right\},
\label{gmod}
\end{eqnarray}
which results in the damping rate:
\begin{equation}
\Gamma\simeq\kappa\left(1-\ln2+{2\langle\mu\rangle T\over m^2_Z}+
{\langle\mu\rangle T^2F_1\over m^3_Z}\right)\Gamma_Z,
\label{coldw}
\end{equation}
with the same geometric factor $F_1$ as in previous case.

iv) Hot, weakly charge-asymmetric fermionic matter, $\mu/T\ll1$,
$m_Z/T\ll1$. Here the above $G$ is
\begin{eqnarray}
G(\mu,|{\bf k}|)&\simeq&{1\over4}|{\bf k}|\left\{|{\bf V}|^2
\left[{\bf k}^2\left(1+{m_Z\mu\over6T^2}\right)+m_Z^2\right]\right.
 \nonumber\\
& &\left.+{\mu{\bf k}^2\over12T^2}\cdot({\bf k}^2-m_Z^2)
i{\bf k}\cdot[{\bf V}\times{\bf V}^\ast]/{\bf k}^2\right\},
\label{gmodh}
\end{eqnarray}
The evaluation of the $CP$-even contribution looks similar to the previous
cases, while the $CP$-odd term requires some care. One has
\begin{eqnarray*}
\int G^{\rm CP-odd}(\mu,|{\bf k}|){d^3k\over(2\pi)^3|{\bf k}|}
&=&{\mu\over12T^2}
\left({4\pi\over\bar g}\right)^2\oint d\sigma_1\oint d\sigma_2
([{\bf X}^\prime_1\times{\bf X}^\prime_2]
\cdot\mbox{\boldmath$\nabla$}_{12})\nonumber\\
& &\times(-1)\int{d^3k\over(2\pi)^3}\theta(m_Z-|{\bf k}|)
\left(1-{3m^2_Z\over{\bf k}^2+m^2_Z}+{2m^4_Z\over({\bf k}^2+m^2_Z)^2}
\right)        \nonumber\\
& &\times\exp(-i{\bf k}\cdot{\bf X}_{12}).
\end{eqnarray*}
When evaluating the integral over $m_Z$-dependent terms, one should have
in mind that the upper integration limit can be set to infinity. Indeed,
the oscillating factor $\sin |{\bf k}||{\bf X}_{12}|$ resulting from
the integration over the polar angle, shows that the most essential
contribution comes from $ |{\bf k}|\sim|{\bf X}_{12}|^{-1}$, which, in
view of inequality $m_Z|{\bf X}_{12}|\gg1$, means that the suggested
approximation is true. All this yields
\begin{eqnarray*}
G^{\rm CP-odd}(\mu,|{\bf k}|)&\simeq&{\mu m^2_Z\over24\pi T^2}
\left({4\pi\over\bar g}\right)^2\oint d\sigma_1\oint d\sigma_2
{({\bf X}_{12}\cdot[{\bf X}^\prime_1\times{\bf X}^\prime_2])
\over|{\bf X}_{12}|}       \nonumber\\
& &\times\left(-{\sin m_Z|{\bf X}_{12}|\over\pi{\bf X}_{12}^2}+
{m^2_Z\over2}\exp(-m_Z|{\bf X}_{12}|)\right).
\end{eqnarray*}
The integral over the first term in the parentheses cannot be further
simplified, but the second one can. Making use of the expansion
Eq. (\ref{expan}) and extending the limits of integration over $z$ to
$\pm\infty$, in view of the fast convergence, one arrives at the
expression for the damping rate:
\begin{equation}
\Gamma\simeq{\kappa\over4}
\left[1-\ln2+{\langle\mu\rangle m_Z\over T^2}(1-{4\over3}\ln2)
+{\langle\mu\rangle m_Z\over3T^2}F_2\right]\Gamma_Z,
\label{hotw}
\end{equation}
with another geometric $P$-odd factor
\begin{equation}
F_2=\oint d\sigma{{\bf X}^\prime\cdot[{\bf X}^{\prime\prime}
\times{\bf X}^{\prime\prime\prime}]\over m^3_ZL}
-\oint d\sigma_1d\sigma_2
{{\bf X}_{12}
\cdot[{\bf X}_1^\prime\times{\bf X}_2^\prime]\over
4\pi Lm_Z{\bf X}_{12}^3}\sin m_Z|{\bf X}_{12}|.
\label{f2}
\end{equation}
Note that for contours with a
smooth dependence on $\sigma$ of the radius of curvature $R$ , the first
term in $F_2$ is
reduced to $$\oint d\sigma{\bf X}^\prime\cdot[{\bf n}\times{\bf n}^\prime
]/m^3_ZLR^2,$$ where ${\bf n}$ is the normal to the contour, and the
integral over $\sigma$ is $2\pi$ times the so called {\it twist} \cite{frankk}
of the contour, while the second term, up to the factor $1/Lm_Z$,
looks similar to the {\it writhing}
number \cite{frankk} (see them in Sec. \ref{sec4}),
if it were not the oscillating factor in the integrand
coming from the nontrivial distribution of Z-magnetic field.

\section{Damping of the basic characteristics of Z-string}
\label{sec5}

Assume that the Z-string with the winding number $n_0$ be formated.
Then its energy is $E_0\simeq\varepsilon_{\rm v}Ln^2_0$.
After that it goes to the state with the  winding number
$n=1$ via either the fermion pair emission, after the time duration
$T_0\simeq2\Gamma^{-1}\ln n_0$, or by some another mechanism mentioned in
\cite{fn2}, whose rate is unknown.
In turn, the state with the single winding number
decays only via the emission of the fermions. Since this state has the energy
$\varepsilon_{\rm v}L$ and is the superposition of the stationary states whose
excitation quantum numbers $N$ are distributed according to the Poisson
formula with the mean $N_0=\varepsilon_{\rm v}L/m_Z$, the number of emitted
fermions is by a factor of $\ln N_0\gg1$ grater than that
in the case of the single  Z$^0$ boson.
Since Z-flux is directly related to the winding number, it decays with the
rate $\Gamma/2$. The energy
depends quadratically on Z-flux, so its damping rate is $\Gamma$.

The energy and Z-flux are nonzero for string configurations of arbitrary
shape, including the straight strings. On the contrary, the helicity,
\begin{eqnarray}
h_Z&=&\int d^3x{\bf Z}\cdot(\mbox{\boldmath$\nabla$}\times{\bf Z})=
\int{d^3k\over(2\pi)^3}
\frac{i{\bf k}\cdot\left[{\bf v}({\bf k},t)
\times{\bf v}^{\ast}({\bf k},t)\right]}{({\bf k}^2m^{-2}_Z+1)^2}
=              \nonumber\\
& &\Biggl(\frac{4\pi}{\bar g}\Biggr)^2\int\frac{d^3k}{(2\pi)^3}
\Biggl(\frac{m^2_Z}{{\bf k}^2+m^2_Z}\Biggr)^2
\sum_{a,b}
\oint\oint d\sigma_a d\sigma_b\exp[-i{\bf k}\cdot({\bf X}_a
-{\bf X}_b)]n_an_b    \nonumber\\
& &\times i{\bf k}\cdot[{\bf X}^\prime_a\times{\bf X}^\prime_b]/{\bf k}^2
\label{hel}                                                     
\end{eqnarray}
\cite{kozhev95b}, is nonzero only for the
configurations of Z-strings which are not invariant under the space inversion.
Indeed, the terms with $a\not=b$, after the momentum integration, give the
linking number,
\begin{eqnarray*}
L[a,b]={1\over4\pi}\oint d\sigma_a\oint d\sigma_b
{{\bf X}_{ab}\cdot[{\bf X}^\prime_a\times{\bf X}^\prime_b]
\over|{\bf X}_{ab}|^3}
\end{eqnarray*}
\cite{frankk}, of two contours, with the exponentially small corrections.
The contribution of the typical term with $a=b$, after the momentum
integration, reads
\begin{eqnarray}
h_Z(a=b)&\propto& W[a]-{1\over4\pi}\oint d\sigma_1\oint d\sigma_2
{{\bf X}_{12}\cdot[{\bf X}^\prime_1\times{\bf X}^\prime_2]\over
|{\bf X}_{12}|^3}
\left(1+m_Z|{\bf X}_{12}|+{1\over2}m^2_Z|{\bf X}_{12}|^2
\right)            \nonumber\\
& &\exp(-m_Z|{\bf X}_{12}|),
\label{ab}
\end{eqnarray}
where ${\bf X}_{12}\equiv{\bf X}_a(\sigma_1)-{\bf X}_a(\sigma_2)$ refers
to the same contour $a$, and
\begin{eqnarray*}
W[a]={1\over4\pi}\oint d\sigma_1\oint d\sigma_2
{{\bf X}_{12}\cdot[{\bf X}^\prime_1\times{\bf X}^\prime_2]
\over|{\bf X}_{12}|^3}
\end{eqnarray*}
\cite{frankk} is the writhing number of the contour $a$. The
$m_Z$-dependent term in Eq. (\ref{ab}) is evaluated with the help of the
expansion (\ref{expan}) to give
$$-{1\over2\pi m^2_Z}\oint d\sigma {\bf X}^\prime_a\cdot
[{\bf X}^{\prime\prime}_a\times{\bf X}^{\prime\prime\prime}_a].$$
In the case of sufficiently smooth contours the latter can be represented as
$-T[a]/(m_ZR)^2$, where
$$T[a]={1\over2\pi}\oint d\sigma{\bf X}^\prime\cdot[{\bf n}\times
{\bf n}^\prime]$$
\cite{frankk} is the twist number of the contour $a$ whose normal vector
is ${\bf n}$ and the radius of curvature is $R$.
So the twist contribution to the helicity is suppressed as
$(Rm_Z)^{-2}$, and the resulting expression for the helicity can be
written as \cite{vachas94,sato95,kozhev95b}
\begin{equation}
h_Z=\left(\frac{4\pi}{\bar g}\right)^2\left\{
\sum_{a}n^2_aW[a]+2\sum_{a<b}n_an_bL[a,b]\right\}.
\label{dop3}
\end{equation}
Since $h_Z$ is just Chern-Simons number of the Z-string field configuration,
it characterizes possible processes with anomalous nonconservation of the
baryon number \cite{rubakov96,vachas94,barrio}.

The rate of the change in the helicity is evaluated semiclassically. To this
end one should first take the expectation value of Eq. (\ref{hel}) in the
quantum  state discussed in Sec. \ref{sec3}.
Further differentiation with respect to time
is then performed similar to the case of the damped classical oscillator,
where the averaging over period is implied, giving the contribution due to
the variable Z-flux to be $\dot h_Z^{\rm Z-flux}=-\Gamma h_Z$. The
contribution coming from the
slow classical contour motion is calculated with the
help of the relation
$${\partial\over\partial t}\oint d\sigma[{\bf k}\times{\bf X}^\prime]
\exp(-i{\bf k}\cdot{\bf X})=
i\oint d\sigma{\bf k}\times({\bf k}\times
[{\bf\dot X}\times{\bf X}^\prime])\exp(-i{\bf k}\cdot{\bf X}),$$
which can be verified by a straightforward calculation. One finds
\begin{eqnarray}
\dot h_Z^{\rm contour}&=&\left({4\pi\over\bar g}\right)^2
\int{d^3k\over(2\pi)^3}\left({m^2_Z\over{\bf k}^2+m^2_Z}\right)^2\sum_{ab}
\langle n_an_b\rangle\oint d\sigma_a\oint d\sigma_b
({\bf\dot X}_a-{\bf\dot X}_b)[{\bf X}^\prime_a
\times {\bf X}^\prime_b]         \nonumber\\
& &\times\exp(-i{\bf k}\cdot{\bf X}_{ab})   \nonumber\\
& &={2\pi m^3_Z\over\bar g^2}\sum_{ab}
\langle n_an_b\rangle\oint d\sigma_a\oint d\sigma_b
({\bf\dot X}_a-{\bf\dot X}_b)[{\bf X}^\prime_a
\times {\bf X}^\prime_b]\exp(-m_Z|{\bf X}_{ab}|).
\label{cont}
\end{eqnarray}
It is clear that the terms with $a\not=b$ give exponentially small
contribution. This is natural, since the analogous terms in the expression
for $h_Z$ give the contribution to the linking number $L[a,b]$ known to be
the topological invariant. The contribution of the terms with $a=b$ is
calculated with the help of Eq. (\ref{expan}). In contrast to the case of
helicity itself, their contribution to the time derivative of the latter
is independent of $m_Z$. In total, one obtains
\begin{equation}
\dot h_Z\simeq-\Gamma h_Z+{8\pi\over\bar g^2}
\sum_a\langle n^2_a\rangle\oint d\sigma{\bf\dot X}_a\cdot[{\bf X}^\prime_a
\times{\bf X}^{\prime\prime\prime}_a].
\label{deriv}
\end{equation}
The second   term in the right hand side
is of purely classical origin. Its
inclusion  is justified in the case when the decay by the fermionic pair
emission is prohibited.

\section{Conclusion}
\label{sec6}

The results obtained in this paper shed a new light on the issue of
stability of the gauge vortex defects in that the latter can be considered as
purely classical field configurations only under the definite circumstances.
The winding number of the scalar field is intimately interrelated with the
gauge field so that if the latter decays the former will do the same, provided
the spectrum of excitations allowed to couple to the string background
satisfies the specific threshold condition. The radial part $f$ (modulus) of
the scalar field is irrelevant in the London limit. The above threshold
condition is satisfied for Z-strings coupled to fermions.
The interaction with fermions
of these extended objects is innate and will result in their decay even in
the models that admits  metastability at the classical level
\cite{garriga,dvali}.
The suppression of this decay could be possible, in principle, at the nonzero
fermionic density in surrounding matter.
The main reason of instability of a sufficiently long Z-string in empty
space is not the collapse (in the case of closed contour), but a much more
fast process of the fermion emission. The numerical estimate for the
critical length $L_0$ is obtained from the condition
$${L_0\over v}\sim{1\over\Gamma}\ln\left({4\pi m_Zv\over\bar g^2\Gamma}
\ln{m_H\over m_Z}\right).$$
Taking the velocity of the shrinkage  $v\sim0.5$, one gets
$L_0m_Z\sim2\times10^3$. The loops with the length greater than
2000 Compton lengths of  Z-boson will evaporate before their shrinkage.


\begin{references}
\bibitem{fn1}
Note that emergence of  $m_Z$ in Eq. (\ref{levels}) as the
frequency of an effective oscillator is the feature of the London limit.
In general, the frequency squared in the case of the straight string
is expressed through the exact profiles of the scalar
and vector fields $f(\rho)$ and $v(\rho)$ possessing the $\rho\to\infty$
asymptotics to be, respectively,  $f\to\eta/\sqrt{2}$ and $v\to-g^{-1}$
as
$$\omega^2=\frac{\int_0^\infty d\rho\rho^{-1}
[v^{\prime2}+2g^2(v+g^{-1})^2f^2]}
{\int_0^\infty d\rho\rho^{-1}(v+g^{-1})^2}.$$ Under the replacement
$f\to\eta/\sqrt{2}$ the above expression gives $m^2_Z$ with the logarithmic
accuracy.
\bibitem{fn2}
There could be another
decay mode of the state with multiple winding number $n>1$
referred usually as the 'classical instability against the decay into the
strings with the unit winding number'.
However, its mechanism has never been specified.
\end{references}
\end{document}